\documentclass[12pt]{article}

\usepackage{epsf,epsfig}
\usepackage{color,colortbl}
\usepackage{graphics}

\def\simgt{\rlap{\lower 3.5 pt \hbox{$\mathchar \sim$}} \raise 1pt \hbox {$>$}}
\def\simlt{\rlap{\lower 3.5 pt \hbox{$\mathchar \sim$}} \raise 1pt \hbox {$<$}}

\def\3half{\textstyle\frac32}

\begin{document}

\begin{titlepage}

\begin{flushright}
{\small
PITHA~05/05\\
hep-ph/0505075\\[0.2cm]
May 5, 2005}
\end{flushright}

\vspace{0.7cm}
\begin{center}
\Large\bf\boldmath
Corrections to $\mbox{sin}(2\beta)$ from CP asymmetries in 
$B^0\to (\pi^0,\rho^0,\eta,\eta^\prime,\omega,\phi) K_S$ decays

\unboldmath
\end{center}

\vspace{0.8cm}
\begin{center}
{\sc M. Beneke}\\
\vspace{0.7cm}
{\sl Institut f\"ur Theoretische Physik E, RWTH Aachen\\
D--52056 Aachen, Germany}
\end{center}

\vspace{1.0cm}
\begin{abstract}
\vspace{0.2cm}\noindent
Neglecting smaller amplitudes the time-dependent CP asymmetry 
in penguin-dominated $b\to s q\bar q$ transitions (such as 
$B\to \phi K_S$) is expected to equal $\pm\sin(2\beta)$, 
an expectation not borne out by the present average experimental 
data. I compute and discuss the correction due to the smaller 
amplitudes in the framework of QCD factorization. 
\end{abstract}

\vfil
\end{titlepage}

\section{Introduction}

The angle $\beta$ of the unitarity triangle has been determined 
to $\sin(2\beta) = 0.725 \pm 0.037$ \cite{Aubert:2004zt} from 
time-dependent CP asymmetries in $b\to c\bar c s$ transitions.  
If sub-leading decay amplitudes can be neglected 
as argued in \cite{Nir:1990mw}, time-dependent 
CP asymmetries in penguin-dominated $b\to s q\bar q$ transitions 
should also take the value $\pm \sin(2\beta)$. There now 
exist various measurements \cite{Chen:2005dr}, which {\em on average} point 
to the significantly smaller value $0.43\pm 0.07$. It is not 
inconceivable that flavour-specific new flavour-violating 
interactions cause anomalous effects in $b\to s$ transitions 
without resulting in inconsistencies with other measurements. This 
would be a rather spectacular resolution of the apparent
discrepancy. But before this conclusion can be drawn, a thorough
study of the sub-leading decay amplitudes is necessary to ascertain 
the Standard Model expectation.  This is undertaken 
here in the framework of QCD factorization \cite{BBNS1}. 

The analysis is based on the next-to-leading order (NLO) 
factorization calculations performed 
in \cite{Beneke:2003zv}, where numerical values of the time-dependent 
CP asymmetries for the $\phi K_S$ and $\eta^\prime K_S$ final states 
have already been given. In this Letter I include 
a larger set of final states (see also the recent 
work \cite{Cheng:2005bg,Buchalla:2005us}), and consider a more detailed error 
estimation that includes a scan of the theoretical parameter 
space \cite{talk}. I also discuss constraints on the sub-leading 
decay amplitudes that do not rely on factorization 
but are inspired by it. Another method to constrain the differences of
time-dependent CP asymmetries in $b\to c\bar c s$ and $b\to s\bar s q$
transitions based on systematic approximations to the strong
interactions relies on the assumption of SU(3) flavour symmetry.  
This results in bounds on the magnitude of this difference, but the 
sign cannot be determined \cite{Grossman:2003qp}. An estimate 
in a model of (long-distance?) final state interactions is 
given in \cite{Cheng:2005bg}.

The time-dependent CP asymmetry in decays to CP eigenstates is given by 
\begin{equation}
\label{cpdef}
\frac{\mbox{Br}(\bar B^0(t)\to f) - \mbox{Br}(B^0(t)\to f)}
 {\mbox{Br}(\bar B^0(t)\to f) + \mbox{Br}(B^0(t)\to f)}
 \equiv  S_f\sin(\Delta m_B\,t) - C_f\cos(\Delta m_B\,t),
\end{equation}
with $\Delta m_B$ the $B^0\bar B^0$ mass difference. The $\bar B$
decay amplitude involves two weak couplings $V_{pb} V^*_{ps}$ and 
two strong interaction amplitudes $a_f^p$. I write
\begin{equation}
\label{ampl}
A(\bar B\to f) = V_{cb} V_{cs}^* \,a_f^c+ V_{ub} V_{us}^* \,a_f^u 
\propto 1 + e^{-i\gamma} \,d_f,
\end{equation}
where 
\begin{equation}
d_f = \epsilon_{\rm KM} \frac{a_f^u}{a_f^c} \equiv 
\epsilon_{\rm KM} \hat d_f
\quad\mbox{with}\quad
\epsilon_{\rm KM} = \left|\frac{V_{ub} V_{us}^*}{V_{cb} V_{cs}^*}\right| 
\sim 0.025.
\end{equation}
A standard calculation now gives
\begin{equation}
\label{dSS}
\Delta S_f \equiv -\eta_f S_f-\sin(2\beta) = 
\frac{2 \,\mbox{Re}(d_f) \cos(2\beta)\sin\gamma +|d_f|^2 
\left(\sin(2\beta+2\gamma)-\sin(2\beta)\right)}
{1 + 2 \,\mbox{Re}(d_f) \cos\gamma + |d_f|^2},
\end{equation}
\begin{equation}
\label{CC}
A_{{\rm CP}, f} \equiv -C_f = \frac{2 \,\mbox{Im} (d_f) \sin\gamma}
{1 + 2 \,\mbox{Re} (d_f) \cos\gamma + |d_f|^2}.
\end{equation}
Here $\eta_f$ denotes the CP eigenvalue of $f$. (All final states
discussed below have $\eta_f=-1$.) The quantity $\Delta S_f$ is the
central object of this Letter. One notes that (a) $d_f$ is suppressed 
by a small ratio of CKM elements, $\epsilon_{\rm KM}$, leading to 
the expectation that $-\eta_f S_f\approx \sin(2\beta)$ (see above); 
(b) if $d_f$ is small as expected, then to first order in $d_f$ 
the two asymmetries $S_f$ and $C_f$ involve {\em independent} hadronic
parameters, namely the dispersive and absorptive part of 
$\hat d_f=a_f^u/a_f^c$. 

\section{\boldmath Anatomy of $\Delta S_f$ in factorization}

The hadronic amplitudes $a_f^p$, $p=u,c$  are sums of 
``topological'' amplitudes, referring to tree ($T, C$), QCD penguin 
($P^p$), singlet penguin ($S^p$), electroweak penguin ($P_{\rm EW}^p, 
P_{{\rm EW},C}^p$) and annihilation contributions.  The relation to 
the ``flavour'' amplitudes used in QCD factorization
\cite{Beneke:2003zv} is $T\leftrightarrow \alpha_1$,  
$C\leftrightarrow \alpha_2$, 
$P^p\leftrightarrow \alpha_4^p+\beta_3^p$, $S^p\leftrightarrow
\alpha_3^p+\beta^p_{S3}$, 
and $(P_{\rm EW}^p, P_{{\rm EW},C}^p) \leftrightarrow 
(\alpha_{3,\rm EW}^p,\alpha_{4,\rm EW}^p)$ with the difference 
that the $\alpha_i$ exclude form factors, decay constants and the 
CKM factors, while the topological amplitudes exclude only 
the CKM factor. In addition, a penguin amplitude such as $P^c$ 
may be a sum of several $\alpha_4^p$ terms depending 
on the flavour flow to the final state. The expressions for all 
relevant decay amplitudes in terms of flavour amplitudes are 
collected in Appendix~A of~\cite{Beneke:2003zv}. Schematically, 
for the strangeness-changing decays $\bar B^0\to M \bar K^0$, 
the hadronic amplitude ratio is given by 
\begin{equation}
\label{schematic}
d_f \sim \epsilon_{\rm KM} \,\frac{\{P^u,C,\ldots\}}{P^c+\ldots},
\end{equation}
where the dominant amplitudes have been indicated. Note that 
the amplitudes $P^u,C,\ldots$ depend on the final state $f$. 

In the QCD factorization framework the topological amplitudes are 
computed in the form \cite{BBNS1}
\begin{eqnarray}
\label{factform}
T,C,P^{c,u},\ldots &=& \sum_{\rm terms} 
C(\mu_h) \times\Big\{F^{BM_1}\times 
\underbrace{T^{\rm I}(\mu_h,\mu_s)}_{1+\alpha_s+\ldots} 
\star f_{M_2}\Phi_{M_2}(\mu_s) 
\nonumber \\
&& \hspace*{-2cm} 
+ \,f_{B}\Phi_B(\mu_s)\star \Big[
\underbrace{T^{\rm II}(\mu_h,\mu_I)}_{1+\ldots}\star 
\underbrace{J^{\rm II}(\mu_I,\mu_s)}_{\alpha_s+\ldots}
\Big] \star f_{M_1}\Phi_{M_1}(\mu_s)\star 
f_{M_2}\Phi_{M_2}(\mu_s)\Big\}
\nonumber \\[0.1cm]
&& \hspace*{-2cm} 
+ \,\mbox{$1/m_b$-suppressed terms}
\end{eqnarray}
reducing the hadronic input to form factors $F^{BM}$ and 
light-cone distribution amplitudes $\Phi_X$. The underbraces indicate 
the order in perturbation theory to which the various short-distance 
kernels are computed at NLO. The numerical implementation of 
(\ref{factform}) also includes some $1/m_b$ power corrections from 
scalar penguin operators, and from an estimate of annihilation 
topologies. The accuracy of the treatment is generically 
limited by $\Lambda_{\rm QCD}/m_b\sim (10-20)\%$ at the amplitude 
level.

The actual uncertainties affect different observables to a different 
degree and must be estimated on a case-by-case basis. The 
``colour-allowed'' amplitudes $T, P^p_{\rm EW}$ are rather certain,
while the ``colour-suppressed'' amplitudes $C, P^p_{{\rm EW}, C}$ 
receive contributions from spectator scattering (the second line 
of (\ref{factform})) enhanced by large Wilson coefficients, and are 
inflicted by larger uncertainties. The QCD penguin amplitudes 
include uncertain annihilation contributions, although the ratio 
$P^u/P^c$ is less affected. Finally, the singlet
amplitude $S^p$ involves several specific decay 
mechanisms \cite{Beneke:2002jn}, which are difficult to compute 
quantitatively, though none of them seems to be of particular
importance for the CP asymmetries. Eq.~(\ref{schematic}) indicates 
that $\Delta S_f$ involves some of the less certain amplitudes. 

The numerical analysis below takes into account all flavour
amplitudes following \cite{Beneke:2003zv}, but it suffices to focus on
a few dominant terms to understand the qualitative features of 
the result. Then, for the various final states, the relevant hadronic 
amplitude ratio is given by
\begin{equation}
\begin{array}{llcll}
\pi^0 K_S \phantom{spa} & 
\hat d_f\sim {\displaystyle \frac{[-P^u]+[C]}{[-P^c]}} & 
\phantom{sp}\phantom{spa} &
\rho^0 K_S  \phantom{spa}& 
\hat d_f\sim {\displaystyle \frac{[P^u]-[C]}{[P^c]}}
\\[0.6cm]
\eta^\prime K_S & 
\hat d_f\sim {\displaystyle \frac{[-P^u]-[C]}{[-P^c]}} & 
\phantom{sp} &
\phi K_S & 
\hat d_f\sim {\displaystyle \frac{[-P^u]}{[-P^c]}} 
\\[0.6cm]
\eta K_S & 
\hat d_f\sim {\displaystyle \frac{[P^u]+[C]}{[P^c]}} & 
\phantom{sp} &
\omega K_S & 
\hat d_f\sim {\displaystyle \frac{[P^u]+[C]}{[P^c]}} 
\end{array}
\end{equation}
The convention here is that quantities in square brackets have 
positive real part. (Recall from (\ref{dSS}) that $\Delta S_f$ 
mainly requires the real part of $\hat d_f$.) In factorization 
$\mbox{Re}\,[P^u/P^c]$ is near unity, 
roughly independent of the particular final 
state, hence $\Delta S_f$ receives a nearly universal, small and 
{\em positive} contribution of about $2 \epsilon_{\rm
  KM}\cos(2\beta)\sin\gamma
\approx 0.03$. On the contrary the magnitudes and 
signs of the penguin amplitudes' real parts can be very different. 
Ignoring uncertainties, I find $|\mbox{Re}\,[P^c]|$ in the 
proportions
\begin{equation}
\begin{array}{ccccccccccc}
\pi^0 K &:& \rho^0 K &:& \eta^\prime K &:& \phi K &:& \eta K &:& \omega K
\\ 
1 &:& 0.5 &:& 2.2 &:& 0.8 &:& 0.5 &:& 0.5
\end{array}
\end{equation}
Hence the influence of the colour-suppressed tree amplitude 
$C$ determines the difference in $\Delta S_f$ 
between the different modes. For 
$(\pi^0, \eta, \omega)K_S$ the effect of $C$ is constructive, 
but for $(\rho, \eta^\prime)K_S$ it is destructive. However, 
the magnitude of $\mbox{Re}\,[P_c]$ is much larger for $ \eta^\prime
K_S$ than for $\rho K_S$, hence $\mbox{Re}\,(\hat d_f)$ remains small 
and positive for the former final state, but becomes negative for 
the latter. 

\section{Factorization results}

\begin{table}[t]
\vspace{0.5cm}
\begin{center}
\begin{tabular}{l|cc|c}
Mode & $\quad \Delta S_f$ (Theory) & 
$\quad\Delta S_f$ [Range]$\quad$ 
& Experiment \cite{Chen:2005dr} (BaBar/Belle)\\
\hline
&&& \\[-0.4cm]
$\pi^0 K_S$
 & $\phantom{-}0.07^{+0.05}_{-0.04}$ & $[+0.02,0.15]$
 & $-0.39^{+0.27}_{-0.29} $ 
   ($-0.38^{+0.30}_{-0.33}$/$-0.43^{+0.60}_{-0.60}$)\\[0.2cm]
$\rho^0 K_S$
 & $-0.08^{+0.08}_{-0.12}$ & $[-0.29,0.02]$
 & ---  \\[0.2cm]
$\eta^\prime K_S$
 & $\phantom{-}0.01^{+0.01}_{-0.01}$ & $[+0.00,0.03]$
 & $-0.30^{+0.11}_{-0.11} $ 
   ($-0.43^{+0.14}_{-0.14}$/$-0.07^{+0.18}_{-0.18}$)\\[0.2cm]
$\eta K_S$
 & $\phantom{-}0.10^{+0.11}_{-0.07}$ & $[-1.67,0.27]$
 & --- \\[0.2cm]
$\phi K_S$
 & $\phantom{-}0.02^{+0.01}_{-0.01}$ & $[+0.01,0.05]$
 & $-0.39^{+0.20}_{-0.20} $  
   ($-0.23^{+0.26}_{-0.25}$/$-0.67^{+0.34}_{-0.34}$)\\[0.2cm]
$\omega K_S$
 & $\phantom{-}0.13^{+0.08}_{-0.08}$ & $[+0.01,0.21]$
 & $-0.18^{+0.30}_{-0.32} $    
   ($-0.23^{+0.34}_{-0.38}$/$+0.02^{+0.65}_{-0.66}$)\\[0.2cm]
\end{tabular}
\end{center}
\caption{\label{tab1} Comparison of theoretical and experimental
  results for $\Delta S_f$.}
\end{table}

The result of the calculation of $\Delta S_f$ is shown in 
Table~\ref{tab1}. The column labeled ``$\Delta S_f$ (Theory)'' 
uses the input parameters (CKM parameters, strong coupling, quark masses, 
form factors, decay constants, moments of light-cone distribution 
amplitudes) summarized in Table~1 of~\cite{Beneke:2003zv}. 
In particular $|V_{ub}/V_{cb}|=0.09\pm 0.02$ and 
$\gamma=(70\pm20)^\circ$ is used. The uncertainty estimate is computed 
by adding in quadrature the individual parameter uncertainties.  
The central values are in good agreement with those given 
in \cite{Cheng:2005bg}, which also uses the input 
from~\cite{Beneke:2003zv}. For the final states $\rho^0 K_S$ and 
$\omega K_S$ they differ from those given in 
\cite{Buchalla:2005us}, where the leading order (naive factorization)
approximation is employed, and the electroweak penguin amplitudes 
are neglected. The next-to-leading order correction 
included in the present calculation 
has a large impact on the branching fractions of penguin-dominated 
modes and is crucial for a successful comparison of QCD factorization 
results with data. Nonetheless, the NLO correction to $\Delta S_f$ is 
never larger than about 30\%, since the amplitude enhancement
partially cancels in the ratio $\hat d_f$. The NLO correction also 
eliminates the large renormalization
scale uncertainty present at leading order. 

The result displays the anticipated pattern. The variation of 
the central value from the nearly universal contribution of
approximately $\epsilon_{\rm KM}$ is due to $\mbox{Re}\,[C/P^c]$, and the 
error comes primarily from this quantity. It is therefore dominated 
by the uncertainty in the hard-spectator scattering contribution 
to $C$, and the penguin annihilation contribution to $P^c$. In 
general one expects the prediction of the asymmetry $S_f$ in 
factorization to be more accurate than the prediction of the 
direct CP asymmetry $C_f$, since $S_f$ is determined by 
$\mbox{Re}\,(a_f^u/a_f^c)$ which is large and calculated at
next-to-leading order, while $C_f$ is determined by 
$\mbox{Im}\,(a_f^u/a_f^c)$, which is small and currently known only at
leading order. The resultant error on $\Delta S_f$ is roughly of the size of 
$\Delta S_f$ itself. Since this is small, one arrives at accurate 
constraints, in particular for the final states $\eta^\prime K_S$ 
and $\phi K_S$. It is striking that the theoretical prediction 
of $\Delta S_f$ is positive, with the exception of $\rho^0 K_S$, 
while the experimental data are all negative.   

Quadratic addition of theoretical errors may not always lead to a
conservative error estimate. Furthermore, the default parameters adopted 
in~\cite{Beneke:2003zv} do not lead to the best description of 
the data. As shown there, a different choice of a few 
parameters (defining certain ``scenarios'') 
results in a very good description of data -- however, some 
observables, in particular the colour-suppressed tree amplitude $C$
important to the present discussion, then take values outside the 
range estimated by quadratic error estimation. To allow for this 
possibility I perform a random scan of the allowed theory parameter 
space. For any observable I take the minimal and maximal value 
attained in this scan to define the predicted range of this 
observable. However, in doing so I discard all theoretical parameter
sets which give CP-averaged branching fractions not compatible 
within 3 sigma with the experimental data, that is I require 
$8.5 < 10^6\,\mbox{Br}\,(\pi^0 K^0)<14.5$,
$0.3 < 10^6\,\mbox{Br}\,(\rho^0 K^0)<9.9$,
$5.3 < 10^6\,\mbox{Br}\,(\phi K^0)<11.9$,
$2.9 < 10^6\,\mbox{Br}\,(\omega K^0)<8.3$,
$10^6\,\mbox{Br}\,(\eta K^0)< 6.0$. No further condition is 
imposed, neither from the corresponding charged decay modes, 
nor any other decay, or from direct CP asymmetries (since these depend 
on other hadronic parameters as mentioned above). Note that 
I also do not require the theoretical parameters to reproduce 
the $\eta^\prime K^0$ branching fraction. The reason for this 
is that in~\cite{Beneke:2003zv} the singlet contribution $F_2$ to 
the $B\to\eta^\prime$ form factor is set to zero simply for 
lack of better information. Since a non-zero $F_2$ can affect 
the branching fraction significantly \cite{Beneke:2002jn}, 
requiring the $\eta^\prime K^0$ branching fraction to reproduce the
data for $F_2=0$ would be overly restrictive on the remaining 
theory parameter space. Nevertheless, one finds that 
the distribution of $B^0\to\eta^\prime K^0$ 
branching fractions generated by the models that survive the 
other branching fraction restrictions has a (broad) maximum 
at $67\cdot 10^{-6}$ in nice agreement with experimental data. 

The resulting ranges for $\Delta S_f$ from a scan of 200000
theoretical parameter sets is shown in the column labeled 
``$\Delta S_f$ [Range]'' in Table~\ref{tab1}. It is seen 
that the ranges are in fact not much different from those 
obtained by adding parameter uncertainties in quadrature -- 
except for the $\eta K_S$ final state, for which almost any 
value of $S_f$ is possible. To understand this exception, one 
must know that similarly large ranges can appear also for 
other final states when no branching fraction restriction is 
imposed. These large values of $\Delta S_f$ originate from 
small regions of the parameter space, where by cancellations 
the leading penguin amplitude $P_c$ becomes very small. This 
leads to large amplifications of $C/P^c$, and 
hence $\Delta S_f$. Such small values of $P^c$ always lead 
to very small branching fractions, hence they are excluded 
by observations except for the case of $\eta K_S$, where no 
lower limit on the branching fraction exists at present. 

\begin{figure}[t]
   \vspace{-0cm}
\hspace*{0.2cm}
   \epsfxsize=7cm
   \epsffile{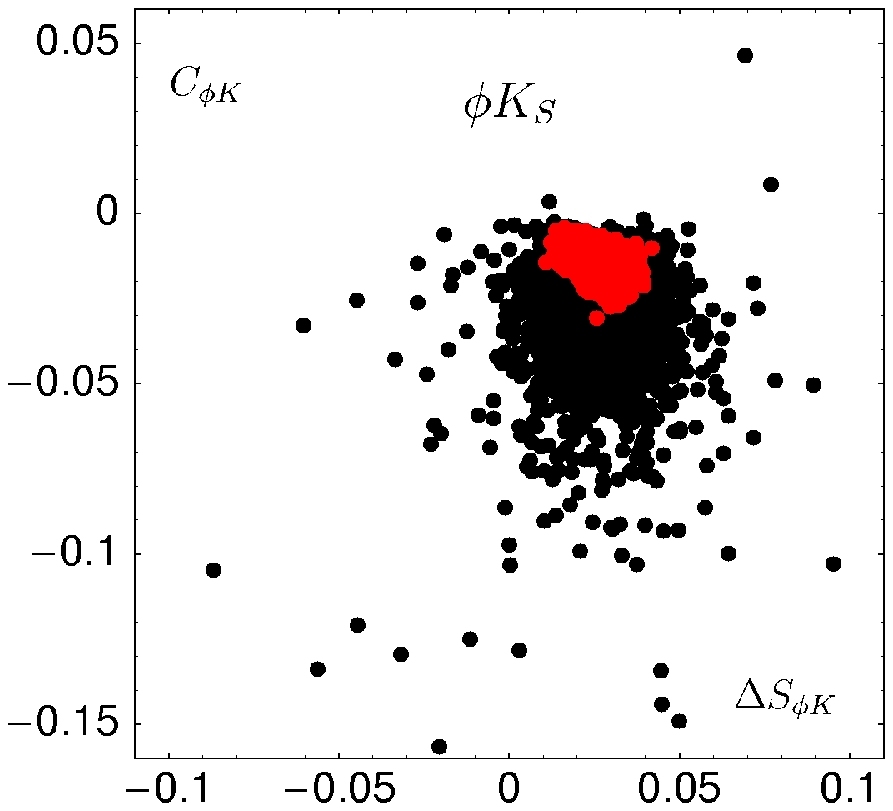}
\hspace*{0.8cm}
   \epsfxsize=6.9cm
   \epsffile{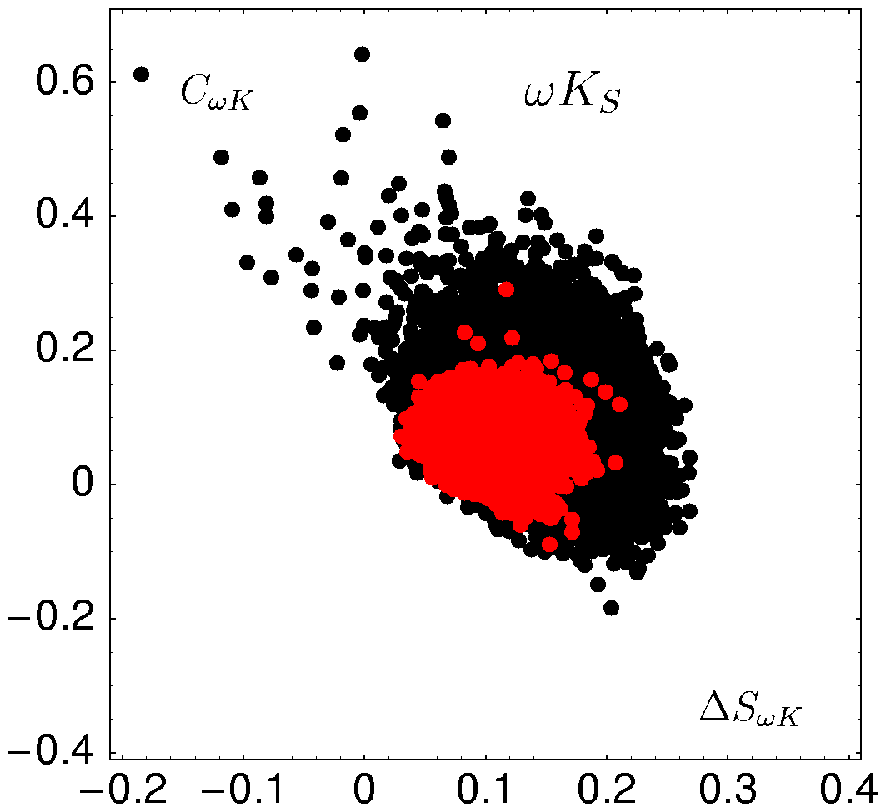}
   \vspace*{-0cm}
\caption{\label{fig1}
Correlation between $\Delta S_f$ and $C_f$ (direct CP asymmetry) 
for $f=\phi K^0$ (left) and $f=\omega K^0$ (right). Theory 
parameter models compatible with the experimental branching 
fractions (as described in the text) are in grey (red), all others in 
black. Based on a sample of 50000 input parameter models.}
\end{figure}

The parameter scan contains more interesting pieces of information 
than the ranges of $\Delta S_f$, since it allows to establish
correlations between $\Delta S_f$ and input parameters, between 
the $\Delta S_f$ for different final states etc. in the framework 
of QCD factorization. For instance, one finds that the ``good'' models
prefer a strange quark mass around $80\,$MeV, smaller renormalization 
scales and a moderate annihilation contribution $\rho_A\approx 
0.7 e^{i\phi_A}$ with $|\phi_A|<70^\circ$, all of which affects 
the magnitude of the dominant QCD penguin amplitude.
Space does not permit a detailed discussion 
here, but Figure~\ref{fig1} shows the correlation between 
$\Delta S_f$ and the direct CP asymmetry $C_f$ 
(see (\ref{cpdef}), (\ref{CC})) taking $f=\phi K^0$ and $\omega K^0$ as
examples. The distribution of points (each corresponding to one theoretical 
parameter set) does not reveal any particular correlation 
between the two observables, especially after the branching fraction 
restriction, as could have been guessed from the 
fact that they mainly involve independent hadronic parameters. 
The Figure also shows that the requirement that the experimental 
branching fractions be reproduced within 3 sigma narrows the 
distribution considerably. Similar conclusions apply to all other 
final states.

\section{Discussion}

Given the important role of $\Delta S_f$ in the detection of anomalous
$b\to s$ flavour transitions, one may question the assumptions that 
go into the factorization approach or attempt to find independent 
validations. Also, given the current experimental status, it would 
already be interesting to know that $\Delta S_f$ should be positive, 
no matter its precise value. Can one establish $\Delta S_f>0$ (except 
for $\rho K_S$) with little assumptions on hadronic physics?

Recall from (\ref{dSS}) that $\Delta S_f$ is roughly 
\begin{equation}
2 \epsilon_{\rm KM}\cos(2\beta)\sin\gamma \,\mbox{Re}
\left(\frac{a^u_f}{a^c_f}\right).
\end{equation}
Large enhancements relative to the factorization predictions 
require an enhancement of the hadronic amplitude ratio. The first
option is a strong suppression of $a^c_f$, but this is excluded 
by the branching fraction measurements (see also the discussion in 
the previous section). The second option is an enhancement  of $a^u_f$ 
by a factor of several. Can this be excluded, or can at least 
the sign of $\mbox{Re}\,(a^u_f/a^c_f)$ be determined?

The only approach to non-leptonic decays other than factorization
based on a small expansion parameter uses SU(3) flavour symmetry 
to relate amplitudes of final states belonging to the same SU(3) 
multiplet. In applications of the method to $\Delta S_f$ one uses 
the branching fractions of $b\to d$ transitions to bound $|d_f|$ 
of the related $b\to s$ transitions \cite{Grossman:2003qp}. 
The best possible limit in 
this method is $|d_f|<\lambda^2\approx 0.05$ ($\lambda$ the Wolfenstein 
parameter), so the {\em theoretical} limit of this method is 
$|\Delta S_f| \,\,\simlt \,\,0.07$. In practice, depending on the values of 
the $b\to d$ branching fractions and the final state $f$, 
the bound is considerably weaker, 
although the region of interesting values (indicated by the
factorization results) may eventually be approached for some 
final states. Note that the sign of $\Delta S_f$ is not determined  
by this method.\footnote{It may be noted that the application of
the SU(3) approach to final states containing $\eta$, $\eta^\prime$, 
$\omega$ and $\phi$ requires additional assumptions beyond 
SU(3). In the SU(3) limit these mesons would be pure octet or 
singlet states, but reality is far from this limit, in particular 
in the case of $\omega$ and $\phi$, 
which are believed to be pure up-down and strange quark states, 
respectively. In \cite{Grossman:2003qp} this SU(3) breaking 
singlet-octet mixing effect is taken into account by assuming 
that the operator matrix elements with the physical meson states are 
related to those with the putative SU(3) states by a single mixing 
angle. This is an assumption that cannot be justified in any 
controlled approximation \cite{Langacker:1974bc}. Rather one must 
introduce a separate mixing angle for every operator. The existence of
large mixing for $\omega-\phi$ and $\eta-\eta^\prime$ should be taken 
as an indication that a SU(3) treatment might be unreliable, since 
for every operator a separate, presumably large, mixing angle must be 
introduced. Phenomenological evidence related to the matrix elements 
of current operators may indicate that 
this SU(3) breaking effect is nearly universal and could be 
described by a single mixing-angle in the 
quark-flavour-basis~\cite{Feldmann:1998vh}, but little is known about 
the matrix elements of the effective weak Hamiltonian.}

A limited amount of information can be obtained from final states 
related to the given one by isospin symmetry, or from 
other observables related to the given final state. 
As already mentioned above, the measurement of 
the direct CP asymmetry ($C_f$) is of limited use if it 
is small, since it constrains the imaginary part of $a^u_f/a^c_f$ 
rather than the real part. On the other hand, a very large direct 
CP asymmetry (for $\phi K_S$, $\eta^\prime K_S$, 
$\pi^0 K_S$) would suggest that $\mbox{Re}(d_f)$ could also be 
large, but this is not rigorous. It would certainly imply large 
violations of factorization, and hence cast doubt on the results 
in Table~\ref{tab1}. No such large direct CP asymmetries have been 
observed to date for the final states discussed here. 

The asymmetry $S_f$ is more closely related to ratios of CP-averaged 
branching fractions, which also depend mainly on real parts of 
amplitude ratios. In the following I consider the pairs 
$(M \bar K^0,M K^-)$, including the charged partners of $M$ 
for $M=\pi,\rho$. The decay amplitudes can be parameterized 
as
\begin{eqnarray}
A(M^- \bar K^0) &=& P+e^{-i\gamma} \,P^u
\nonumber\\
\sqrt{2}\,A(M^0 K^-) &=& [P+P^{EW}]+e^{-i\gamma} \,[T+C+P^u]
\nonumber\\
A(M^+ K^-) &=& [P+P^{C,EW}]+e^{-i\gamma} \,[T+P^u]
\nonumber\\
\sqrt{2}\,A(M^0 \bar K^0) &=&
[-P+P^{EW}-P^{C,EW}]+e^{-i\gamma}\,[C-P^u]
\label{first}
\end{eqnarray}
for $M=\pi, \rho$ (assuming isospin symmetry), and 
\begin{eqnarray}
A(M K^-) &=& [P+P^{C,EW}]+e^{-i\gamma} \,[T+C+P^u]
\nonumber\\
A(M \bar K^0) &=& P+e^{-i\gamma}\,[C+P^u]
\label{second}
\end{eqnarray}
for the remaining $M=\eta^{(\prime)},\phi,\omega$. The notation is 
chosen so that it indicates the dominant contribution to each 
amplitude; the dependence of $P$, $P^u$, ... on $M$ is not spelled
out. In (\ref{second}) the CKM-suppressed penguin amplitude 
$P^u$ is redundant and could be absorbed into $C$. For $M=\phi$ 
the ``tree'' amplitudes $T,C$ are actually annihilation amplitudes and
thus very small, provided $\phi$ is a pure $s\bar s$
state, as will be assumed here. It is 
clear from (\ref{second}) that nothing can be learned 
from the charged decay for  $M=\eta^{(\prime)},\phi,\omega$ without
additional assumptions, since it involves two new 
amplitudes (the colour-suppressed electroweak penguin $P^{C,EW}$, 
and $T$). However, I shall now expand the ratios of CP-averaged 
branching fractions under the premise that certain amplitude ratios 
are small. To this end, note that $T,C,P^u$ which multiply
$e^{-i\gamma}$ are proportional to $\epsilon_{\rm KM}$, while 
the electroweak penguin amplitudes are suppressed by the
electromagnetic coupling. Defining $x\equiv X/P$ and counting 
$\epsilon_{\rm KM}\sim \lambda^2$ with $\lambda$ a counting 
parameter of order $1/5$, the natural magnitudes of the amplitude 
ratios are $t, p^{EW}\sim \lambda$, and $c,p^u,p^{C,EW}\sim
\lambda^2$. Estimates of the real parts of the amplitude ratios are 
given in Table~\ref{tab2} using the scenario IV of
\cite{Beneke:2003zv} as input. In the following discussion, 
$c$ and $p^u$ are allowed to be enhanced to order $\lambda$.

\begin{table}[t]
\vspace{0.5cm}
\begin{center}
\begin{tabular}{l|ccccc}
Modes & $t$ & $c$ & $p^u$ & $p^{EW}$ & $p^{C,EW}$ \\
\hline
&&&&& \\[-0.4cm]
$\pi K$ & $-0.13$ & $-0.06$ & $0.02$ 
        & $\phantom{-}0.13$ & $\phantom{-}0.03$ \\[0.2cm]
$\rho K$ & $\phantom{-}0.27$ & $\phantom{-}0.13$ & $0.01$ 
         & $-0.29$ & $-0.07$ \\[0.2cm]
$\eta^\prime K$ & $-0.03$ & $-0.01$ & $0.02$ 
         & \phantom{-}--- & $\phantom{-}0.01$ \\[0.2cm]
$\eta K$ & $\phantom{-}0.34$ & $\phantom{-}0.14$ & $0.02$ 
         & \phantom{-}--- & $-0.05$ \\[0.2cm]
$\phi K$ & $\phantom{-}0.01$ & $\phantom{-}0.00$ & $0.02$ 
         & \phantom{-}--- & $\phantom{-}0.01$ \\[0.2cm]
$\omega K$ & $\phantom{-}0.23$ & $\phantom{-}0.11$ & $0.02$ 
           & \phantom{-}--- & $-0.08$ \\[0.2cm]
\end{tabular}
\end{center}
\caption{\label{tab2} Estimates of the {\em real part} of the 
amplitude ratios $x$ in scenario IV of \cite{Beneke:2003zv}.}
\end{table}

Turning first to $f=\eta^{(\prime)}K,\phi K,\omega K$,
Eq.~(\ref{second}) implies 
that even when an enhancement of the amplitudes $c,p^u$ by a factor of 
several to order $\lambda$ is allowed, they do not 
appear in the ratio of CP-averaged branching fractions at first order 
in $\lambda$. Thus, with an accuracy of a few percent, 
\begin{equation}
R(f)\equiv \frac{\tau_{B^0} \,\mbox{Br}\,(M^0 K^-)}
{\tau_{B^+} \,\mbox{Br}\,(M^0 \bar K^0)} \approx
1+2\cos\gamma \,\mbox{Re}\,(t).
\end{equation}
Hence $\mbox{Re}\,(t)$ can be determined from data, if $R(f)$ 
is sufficiently different from 1 (to justify the neglect of the 
order $\lambda^2$ terms), but   
$\Delta S_f \propto \mbox{Re}\,(c+p^u)$. The colour-allowed 
tree amplitude $T$ is believed to be well-predicted in factorization, 
and has a small absorptive part. Assuming this, an accurate 
measurement of $R(f)$ for $f=\eta^{(\prime)}K, \omega K$ provides 
an estimate of  $\mbox{Re}\,(P)$, of which the sign should be 
reliable. Making the same assumption for $C$ constrains the
contribution from  $\mbox{Re}\,(c)$ to  $\Delta S_f$, but in this case
the assumption is already questionable. The contribution from 
$\mbox{Re}\,(p^u)$ is not constrained as long as it is of order 
$\lambda$. However, one may argue that {\em if} $\mbox{Re}\,(p^u)$ 
is enhanced to order $\lambda$ by whatever mechanism, 
then -- {\em probably} --  the absorptive part 
$\mbox{Im}\,(p^u)$, and hence the direct CP asymmetry, 
will also be of order $\lambda$. Similar arguments can be 
applied to the $\pi K$ and $\rho K$ system (\ref{first}). To linear
order in $\lambda$, 
\begin{equation}
R(f) \approx \left|\frac{1+p^{EW}}{1-p^{EW}}\right|^2\,
\Big(1+2\cos\gamma \,\mbox{Re}\,(t+2 c)\Big).
\end{equation}
The electroweak penguin amplitudes are now important. For 
$\rho K$ the corresponding prefactor reduces the branching fraction 
ratio by a factor of three. In fact, the contribution is so 
large that the linear approximation becomes inapplicable to the 
$\rho K$ final state. For $\pi K$, the complete set of 
three branching fraction 
ratios can be used in principle to determine the real parts of $t$, $c$ and 
$p^{EW}$ simultaneously with a relative uncertainty of order 
$\lambda$ in the linear approximation. However, the current 
experimental $\pi K$ data does not lead to useful results. 

I conclude from this discussion that it is very difficult to constrain 
$\Delta S_f$ independent of theoretical assumptions using only 
experimental data (other than the measurement of $\Delta S_f$
itself). With some plausible
dynamical assumptions bounds can be derived using SU(3), or 
the real parts and signs of amplitudes related to the quantities 
of interest can be determined and compared to the factorization
calculations, thus providing cross-checks.  

\section{Conclusion}

QCD factorization calculations of the time-dependent CP asymmetry 
in hadronic $b\to s$ transitions yield only small corrections 
to the expectation $-\eta_f S_f \approx \sin(2\beta)$. With 
the exception of the $\rho^0 K_S$ final state the correction 
$\Delta S_f$ is positive, slightly strengthening the discrepancy 
with the current average experimental data. The effect and theoretical 
uncertainty is particularly small for the two final states 
$\phi K_S$ and $\eta^\prime K_S$ already 
analyzed in \cite{Beneke:2003zv}; the calculation of 
$\Delta S_f$ for the final states $\rho^0 K_S$ and $\eta K_S$,
however, is more susceptible to errors because of amplitude 
cancellations. The final-state dependence of $\Delta S_f$ is 
ascribed to the colour-suppressed tree amplitude.

It appears difficult to constrain $\Delta S_f$ 
theory-independently by other observables. In particular, the 
direct CP asymmetries or the charged decays corresponding to 
$f=M K_S$ probe hadronic quantities other than those relevant 
to $\Delta S_f$, 
if these observables take values in the expected range. Large
deviations from expectations such as large direct CP 
asymmetries would clearly indicate a defect in our understanding of
hadronic physics, but even then the quantitative implications for 
$S_f$ would be unclear. 
A hadronic interpretation of large 
$\Delta S_f$ would probably involve an unknown long-distance 
effect that discriminates strongly between the up- and charm-penguin 
amplitude resulting in an enhancement of the up-penguin 
amplitude. No model is known to me that could plausibly produce 
such an effect.

\subsubsection*{Acknowledgement}

I would like to thank M.~Neubert for collaboration 
on~\cite{Beneke:2003zv}, which forms the basis of this analysis, 
and S.~J\"ager for careful reading of the manuscript. 
This work is supported by the DFG Sonder\-forschungsbereich/Transregio~9 
``Computergest\"utzte Theoretische Teilchenphysik''.


\end{document}